\documentclass[]{jfm}

\usepackage{natbib}
\usepackage{graphicx}
\usepackage{amsmath}
\usepackage{array}

\usepackage[utf8]{inputenc}
\usepackage{booktabs}

\newcommand{\aver}[1]{ \! \left\langle {#1} \right \rangle \!}
\newcommand{\vect}[1]{\mathbf{#1}}
\newcommand{\ud}{\mbox{d}}

\title[]{Turbulent drag reduction over curved walls}

\begin{document}

\author[J.Banchetti, P.Luchini \& M.Quadrio]{
J\ls A\ls C\ls O\ls P\ls O\ns B\ls A\ls N\ls C\ls H\ls E\ls T\ls T\ls I$^1$, \ls
P\ls A\ls O\ls L\ls O\ns L\ls U\ls C\ls H\ls I\ls N\ls I$^2$ \ls
\and \ns
M\ls A\ls U\ls R\ls I\ls Z\ls I\ls O\ls \ns Q\ls U\ls A\ls D\ls R\ls I\ls O$^1$
}
\affiliation{
$^1$ Dipartimento di Scienze e Tecnologie Aerospaziali, Politecnico di Milano,
via La Masa 34, 20156 Milano, Italy
\\ [\affilskip]
$^2$ Dipartimento di Ingegneria Industriale, Universit\`a di Salerno, 84084 Fisciano, Italy 
}

\date{\today}

\maketitle

%---------------
\begin{abstract}
This work studies the effects of skin-friction drag reduction in a turbulent flow over a curved wall, with a view to understanding the relationship between the reduction of friction and changes to the total aerodynamic drag. Direct numerical simulations (DNS) are carried out for an incompressible turbulent flow in a channel where one wall has a small bump; two bump geometries are considered, that produce mildly separated and attached flows. Friction drag reduction is achieved by applying streamwise-travelling waves of spanwise velocity (StTW). 

The local friction reduction produced by the StTW is found to vary along the curved wall, leading to a global friction reduction that, for the cases studied, is up to 10\% larger than that obtained in the plane-wall case. Moreover, the modified skin friction induces non-negligible changes of pressure drag, which is favorably affected by StTW and globally reduces by up to 10\%. The net power saving, accounting for the power required to create the StTW, is positive and, for the cases studied, is one half larger than the net saving of the planar case. The study suggests that reducing friction at the surface of a body of complex shape induces further effects, a simplistic evaluation of which might lead to underestimating the total drag reduction. 
\end{abstract}

%----------------------
\section{Introduction}
Flow control aimed at reducing the skin-friction drag on a solid body immersed in a moving fluid is an active research area, motivated by its potential for significant energy savings and reduced emissions in the transport sector. Techniques for turbulent skin-friction drag reduction span from simple passive strategies to active approaches. The present work focuses on the latter group, since it generally produces larger effects which are easier to identify, and in particular considers the streamwise traveling waves (StTW) of spanwise wall forcing, introduced by \cite{quadrio-ricco-viotti-2009}, a technique capable to deliver substantial net savings.

The existing proofs of concept for skin-friction drag reduction are mostly limited to (i) low-Reynolds-number turbulent flows, and (ii) elementary geometries, such as flat plates and straight ducts. One naturally wonders whether the established benefits scale up when limitations (i) and (ii) are relaxed. Recently, limitation (i) has been shown not to hinder large drag reductions by spanwise forcing at high $Re$. For example \cite{gatti-quadrio-2016} estimated that a skin-friction reduction of around 23\% is still possible with moderate-amplitude StTW at flight Reynolds number. 

Owing to issue (ii), though, how to assess drag reduction in practical applications, often characterized by curved walls and/or non-uniform pressure gradients, remains an interesting open problem. For example, \cite{atzori-etal-2018}  recently applied drag reduction (via uniform blowing and body-force damping of near-wall turbulent fluctuations) to a finite wing slab studied by DNS/LES. Because of the complexity of the flow, however, the influence of curvature on the drag reduction effectiveness could not be singled out, as concurrent flow phenomena (like transition and separation) prevent a direct and quantitative comparison with flat-plate boundary layer or plane channel flow. Moreover, since transition was obtained by tripping the flow shortly downstream of the leading edge, the actuation was applied in regions where the wall is almost flat. 

The present work aims at understanding the interaction between skin-friction reduction (produced by StTW) and the overall aerodynamic drag in the simpler setting of a channel flow, where one wall has a bump. The turbulent flow over a plane wall with a bump has been considered several times in the past, both experimentally and numerically, as a representative case of wall-bounded flow with localized wall curvature. The experimental work by \cite{almeida-etal-1993} evolved into the ERCOFTAC C.18 and C.81 test cases, dealing with the flow over two-dimensional periodic hills (polynomial-shaped obstacles) with recirculation in their wake \citep{temmerman-leschziner-2001}. Over the years, such geometry has been employed for validation of various numerical methods, LES subgrid models, and RANS simulations. \cite{breuer-etal-2009} successfully compared the experimental information with results from two different DNS, one of which using the immersed-boundary method, and explored the effect of the Reynolds number. A periodic-hill experiment was designed by \cite{rapp-manhart-2011} to reproduce the configuration often used in numerical simulations, and \cite{khaler-etal-2016} used the same setup with high-resolution particle-image and particle-tracking velocimetry. Their results emphasized the importance of adequate near-wall spatial resolution in the surroundings of the bump. \cite{wu-squires-1998} studied with LES the adverse-pressure-gradient boundary layer created by a bump with a circular arc shape, in an attempt to reproduce the previous experimental study by \cite{webster-degraaff-eaton-1996}. Their results showed that a coarse LES does not provide an entirely accurate description of the experimentally observed small-scale vortical structures in the near-wall region. 

\cite{marquillie-laval-dolganov-2008}, inspired by \cite{bernard-etal-2003}, designed a bump with a fore/aft asymmetry to qualitatively resemble an airfoil. They studied via DNS the budget equations for turbulent kinetic energy. A strong blockage is present in their case, where the flow almost separates over the upper flat wall; a long streamwise distance is required to recover the undisturbed conditions downstream of the bump. In a follow-up study, \cite{marquillie-ehrenstein-laval-2011} increased the value of the Reynolds number and extended the analysis to the vorticity and streaks dynamics, discussing the role of near-wall streaks in the kinetic energy production. More recently \cite{mollicone-etal-2017, mollicone-etal-2018} returned to the arc-shaped symmetric bump to numerically study the process of turbulent separation. Different bulge geometries and Reynolds numbers were considered, and the production, transfer and dissipation of turbulent kinetic energy were analysed via the generalized Kolmogorov equation. In \cite{mollicone-etal-2019} a similar setup, with a smooth bump defined by a cosine function, was used to study particle-laden flows at finite values of the Stokes number. 

The bump flow has also been used to investigate via numerical simulations the effectiveness of flow control applied over complex geometries. In particular, separation control has been addressed by \cite{fournier-etal-2010} via pulsed and continuous jets, and by \cite{yakeno-etal-2015} via plasma actuators. Active flow control is the background of the present study too. Aided by the simplicity of the bumped-wall geometry in the confined setting of a channel flow, we aim at understanding how the skin-friction drag reduction enforced by StTW alters the turbulent flow, its global aerodynamic loads and the power budget. The paper begins with the description of the bump geometry, the numerical method and the simulation parameters in \S\ref{sec:simulations}. Instantaneous and mean flow fields are described in \S\ref{sec:fields}, whereas \S\ref{sec:distributions} contains a quantitative analysis of friction and wall pressure distributions, providing drag coefficients for distributed and concentrated losses. The global power budget is studied in \S\ref{sec:budget}, and \S\ref{sec:conclusions} contains a concluding discussion.

\section{Simulations}
\label{sec:simulations}

Our work deals with a non-planar incompressible turbulent channel flow, studied via Direct Numerical Simulations (DNS). One of the channel walls is flat, and the other has a relatively small two-dimensional bump. Two bump profiles with the same height are considered, to produce an attached and a separated flow. Streamwise-travelling waves for the reduction of frictional drag are imposed at the lower non-planar wall, and their effect on the total drag is measured.

The DNS code, introduced by \cite{luchini-2016}, solves the incompressible Navier--Stokes equations written in primitive variables on a staggered Cartesian grid. Second-order finite differences are used in every direction. The momentum equations are advanced in time by a fractional time-stepping method using a third-order Runge-Kutta scheme. The Poisson equation for the pressure is solved by an iterative SOR algorithm. The non-planar wall is dealt with via an implicit immersed-boundary method, implemented in staggered variables to be continuous with respect to boundary crossing and numerically stable at all distances from the boundary \citep{luchini-2013, luchini-2016}.

\begin{figure}
\includegraphics[width=\textwidth]{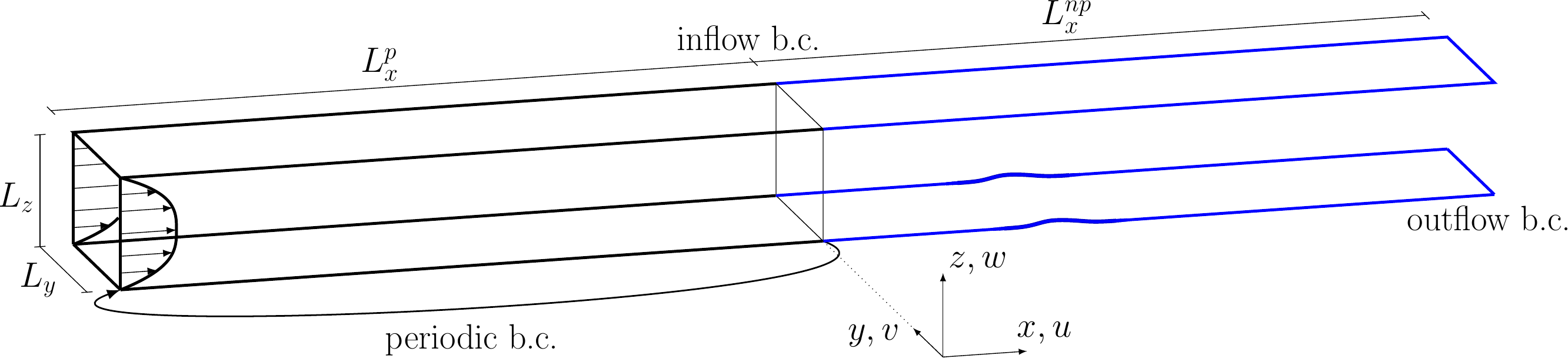}
\caption{Sketch of the computational domain and the reference system. The bump is on the lower wall. The streamwise-periodic upstream domain (black) provides an inflow condition for the downstream one (blue).}
\label{fig:sketch}
\end{figure}

The computational domain (a sketch is shown in figure \ref{fig:sketch}, with the bump on the lower wall) is made by two streamwise-adjacent portions of similar length: the upstream volume with planar walls is streamwise-periodic, and feeds the downstream one where inflow and outflow conditions are used. Periodic conditions are used everywhere for the spanwise direction, and no-slip and no-penetration are enforced on the walls. The outflow condition extrapolates the velocity components according to:
\begin{equation}\label{eq:outlet}
\frac{\partial u_i}{\partial t}+ U_c \frac{\partial u_i}{\partial x} = 0 , \qquad i=1,2,3
\end{equation}
where $U_c(z)$ is the profile of the mean convection velocity of turbulent fluctuations defined as in \cite{quadrio-luchini-2003}, and implemented as the mean velocity profile of the plane channel flow, modified in the near-wall region to have $U_c^+(z) \geq 10$. Alternative outflow conditions have been tested, finding negligible differences in the results.

The simulations are carried out at a bulk Reynolds number $Re_b = U_b h / \nu = 3173$ which in the reference case corresponds to a friction Reynolds number of $Re_\tau = u_\tau h / \nu= 200$ in the plane channel. In their definition, the length scale is $h$, half the distance between the plane walls, whereas the velocity scale is the bulk velocity $U_b$ in the former case and the friction velocity $u_\tau$ in the latter. Unless otherwise noted (e.g. with the plus notation indicating viscous units), in the following, quantities are made dimensionless with $h$ and $U_b$. 

The size of the computational domain is $(L_x^p+L_x^{np}, L_y, L_z) =(24.56, \pi, 2)$ in the streamwise, spanwise and wall-normal directions respectively. The flat walls are placed at $z=0$ and $z=2$. The upstream periodic portion, with streamwise length of $L_x^p = 4 \pi$, runs a standard channel flow DNS where the Constant Flow Rate (CFR) condition is imposed \citep{quadrio-frohnapfel-hasegawa-2016}, and has a spatial resolution of $(n_x, n_y, n_z) =(360,312,241)$ discretization points. The downstream portion of the computational domain starts at $x=0$ with a length of $L_x^{np} = 12$, over which 800 discretization points are non-uniformly distributed; grid and domain sizes in the spanwise and wall-normal directions are the same of the upstream domain, to avoid interpolation.

The grid is tuned for optimal use of computational resources while providing the necessary spatial resolution and smooth description of the bump geometry via the immersed-boundary method. The spanwise grid spacing is uniform at $\Delta y = 0.01$; it corresponds to $\Delta y^+ =2$ based on the inlet $u_\tau$, and to $\Delta y^+ =4$ close to the bump tip, where friction velocity is maximum. Streamwise resolution is uniform at $\Delta x = 0.04$ or $\Delta x ^+ =8$ in the periodic part, but increases as the bump is approached, reaching up to $\Delta x^+= 2$ (based on local $u_\tau$). The wall-normal spacing is neither constant in $z$ nor symmetrical with respect to the centerline, since the bump is present on one wall only. A constant $\Delta z = 0.001$ is adopted from the lower wall to $z=h_b$, where $h_b$ is the maximum bump height, and corresponds to $\Delta z^+= 0.2$, based on the inlet $u_\tau$. $\Delta z$ then gradually increases until, at the centerline, the maximum value of $\Delta z = 0.02 $ is reached. The spacing then decreases again in the upper half of the channel, to reach $ \Delta z = 0.004$ at the upper wall. Overall, the largest streamwise spacing is 6 times the local Kolmogorov length $\eta$, and the wall-normal and spanwise spacings are everywhere less than $2 \eta$. Near the bump the resolution is even higher; in the recirculation zone, the smallest dissipative scales are well resolved, with spacing in every direction equal or lower than $\eta$.

\begin{figure}
\centering
\includegraphics[width=\textwidth]{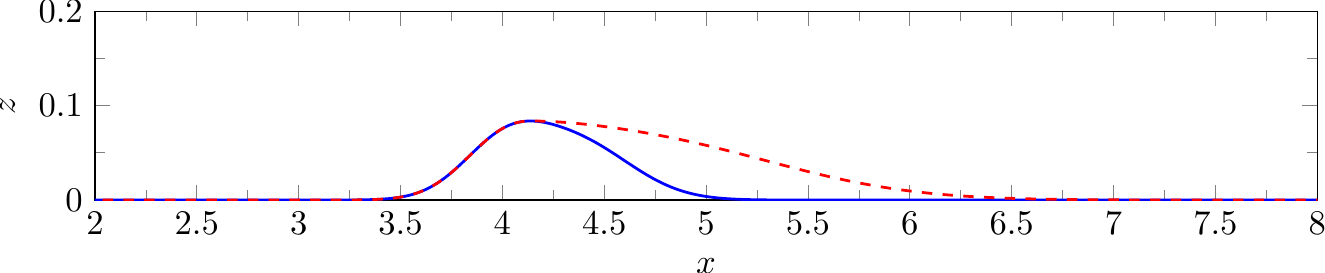}
\caption{Bump geometries $G_1$ (blue line) and $G_2$ (red dashed line); they are identical up to the bump tip. Both have height of $h_b =0.0837$ (only a portion of the streamwise extent is shown; note the enlarged vertical axis). $G_1$ leads to a mildly separated flow, while $G_2$ produces an attached flow.}
\label{fig:geom}
\end{figure}

The geometry of the bump, which is located on the lower wall, is two-dimensional and similar to the one considered by \cite{marquillie-laval-dolganov-2008}, but with significantly smaller size, to reduce blockage and produce nearly undisturbed flow at the inlet and outlet sections. To enable reproducibility, the bump is analytically specified as the sum of two overlapping Gaussian curves, resulting in a smooth profile described by six parameters:
\begin{equation}
G_1(x) = a \exp \left[ - \left( \frac{x - b}{c}  \right)^2 \right] + 
         a'\exp \left[ - \left( \frac{x - b'}{c'} \right)^2 \right] .
\label{eq:bump-definition}
\end{equation}

The parameters values chosen for the geometry $G_1$ are $a = 0.0505, b = 4, c = 0.2922$ and $a' = 0.060425, b' = 4.36, c' = 0.3847$; they produce a bump with height $h_b = 0.0837$. A second geometry $G_2$ is identical to $G_1$ in the fore part up to the tip, but a streamwise expansion factor of 2.5 is applied to the rear part.  Both $G_1$ and $G_2$ are shown in figure \ref{fig:geom}, with the former geometry producing a mildly separated flow, and the latter a fully attached flow. 

In terms of computational procedures, after reaching statistical equilibrium flow statistics are accumulated over a simulation time of $T =1000$. The time step is set at $\Delta t = 1.5 \cdot \ 10^{-3}$, corresponding to an average CFL number of approximately $0.5$. Simulations are carried out with and without StTW for the reduction of skin friction. This specific drag reduction technique, introduced by \cite{quadrio-ricco-viotti-2009}, has been selected because of its interesting energetic properties, its large effect on the turbulent friction and the availability of successful experimental implementations starting from \cite{auteri-etal-2010}. It should also be mentioned that a preliminary version of this work employed another spanwise forcing technique made by stationary waves, and the main findings were the same. StTW are applied at the lower wall only, but including the periodic upstream part. The forcing translates into a non-homogeneous boundary condition for the spanwise velocity component at the wall as follows: 
\begin{equation}
V_w(x,t) = A \sin \left( \kappa_x x - \omega t \right) .
\end{equation}
where $V_w$ is the spanwise velocity at the wall, $A$ is its maximum amplitude, and $\kappa_x$ and $\omega$ represent the spatial and temporal frequencies of the wave. The wall forcing produces a sinusoidal distribution of spanwise velocity which travels in the streamwise direction. The numerical values of the forcing parameters are chosen, based on existing information \citep[e.g.][]{gatti-quadrio-2016}, to guarantee large amounts of skin-friction drag reduction in the plane channel. The selected values $A=0.75$, $\omega=\pi/10$ and $\kappa_x = 2$ yield 46\% of drag reduction in a plane channel at $Re_\tau=200$. Owing to the rather large actuator intensity, the total power budget is only mildly positive, with 11\% of net power savings.

%%%%%%%%%%%%%%%%%%%%%%%%%%%%%%%%%%%%%%%%%%%%%%%
\section{Instantaneous and mean flow fields}

\label{sec:fields}
\begin{figure}
\centering
\includegraphics[width=\textwidth]{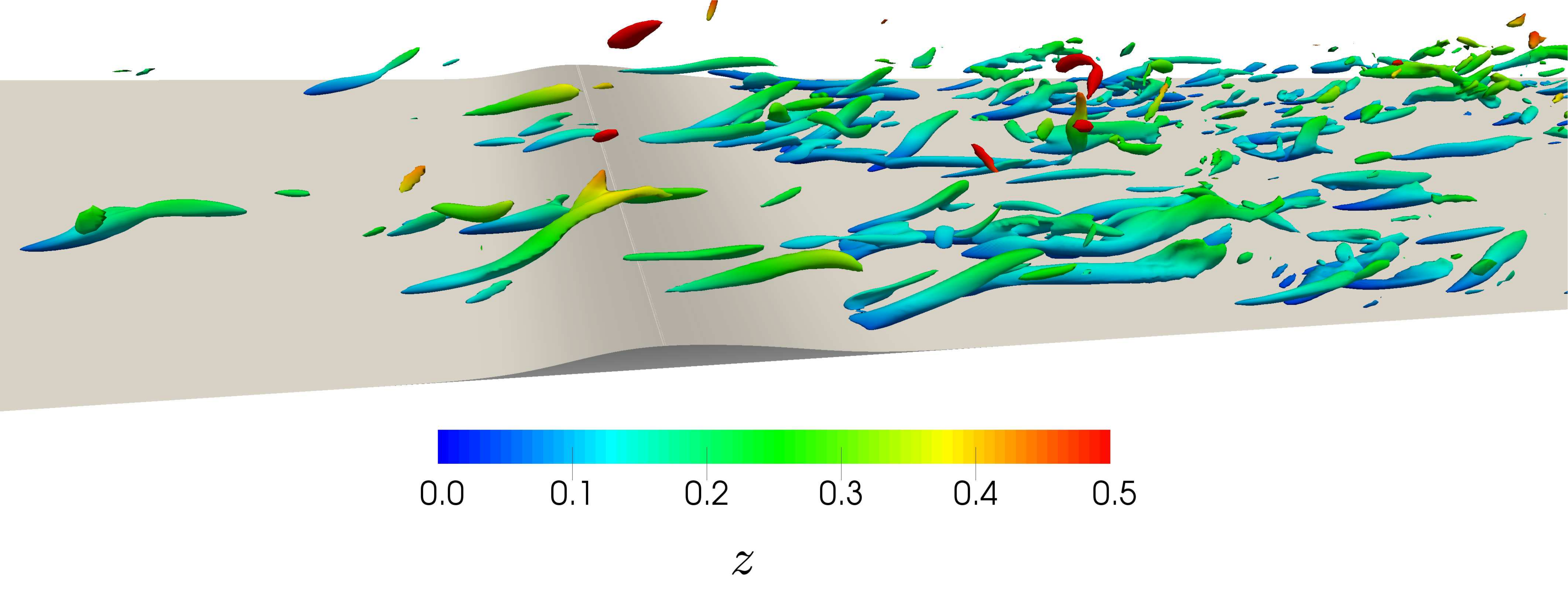}
\caption{Isosurfaces of $\lambda_2^+=-0.04$ for an instantaneous flow field in the reference case. Isosurfaces are colour-coded with the coordinate $z$.}
\label{fig:l2}
\end{figure}

To begin with a qualitative picture of the flow, figure \ref{fig:l2} portraits the appearance of turbulent vortical structures over the shorter bump $G_1$, that will be the focus of this section. The figure is for the reference simulation without StTW, and plots isosurfaces of the intermediate eigenvalue $\lambda_2$  of the velocity gradient tensor \citep{jeong-hussain-1995}, colour-coded with the coordinate $z$. Even though the  height of the bump is quite limited, the localized increase of turbulent activity immediately downstream of the bump is readily appreciated.

\begin{figure}
\centering
\includegraphics[width=\textwidth]{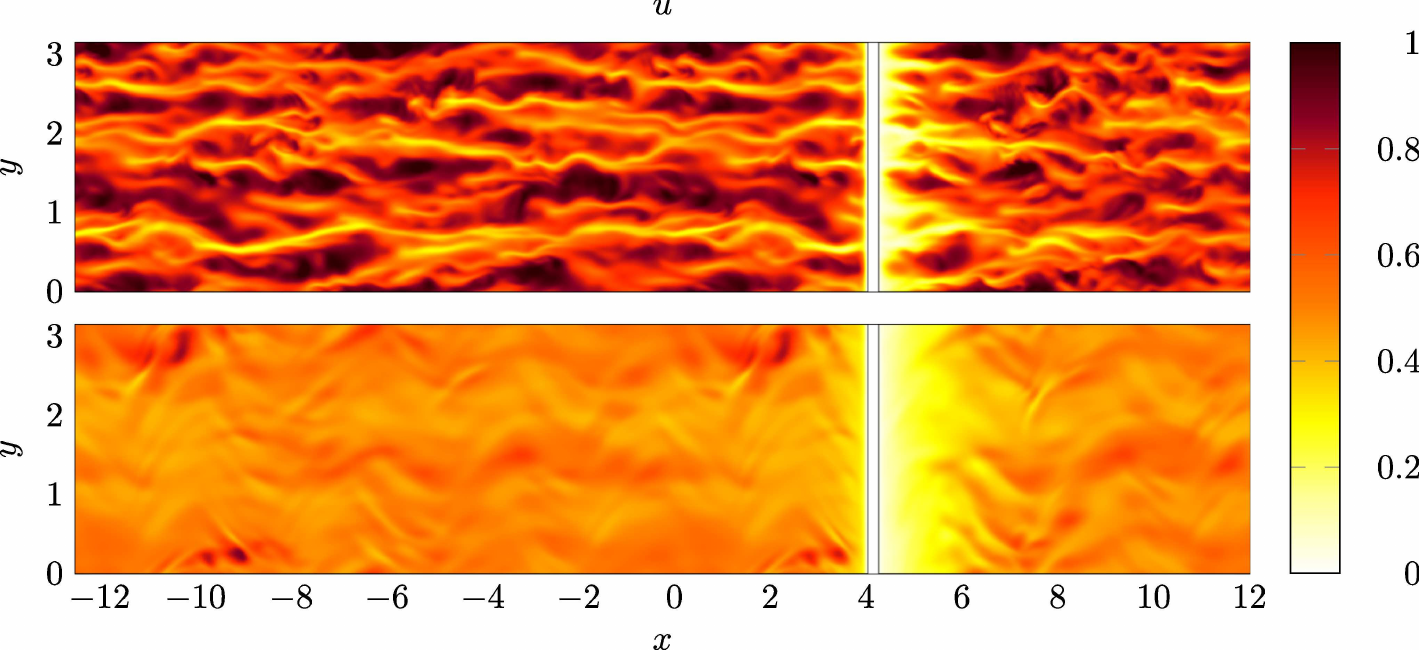}
\caption{Colour plot of an instantaneous streamwise velocity field, in the plane $z = 0.08$ over the bump $G_1$, for the reference case (top) and with StTW (bottom). Flow is from left to right, and the upstream periodic section ends at $x=0$.}
\label{fig:sez-u}
\end{figure}

\begin{figure}
\centering
\includegraphics[width=\textwidth]{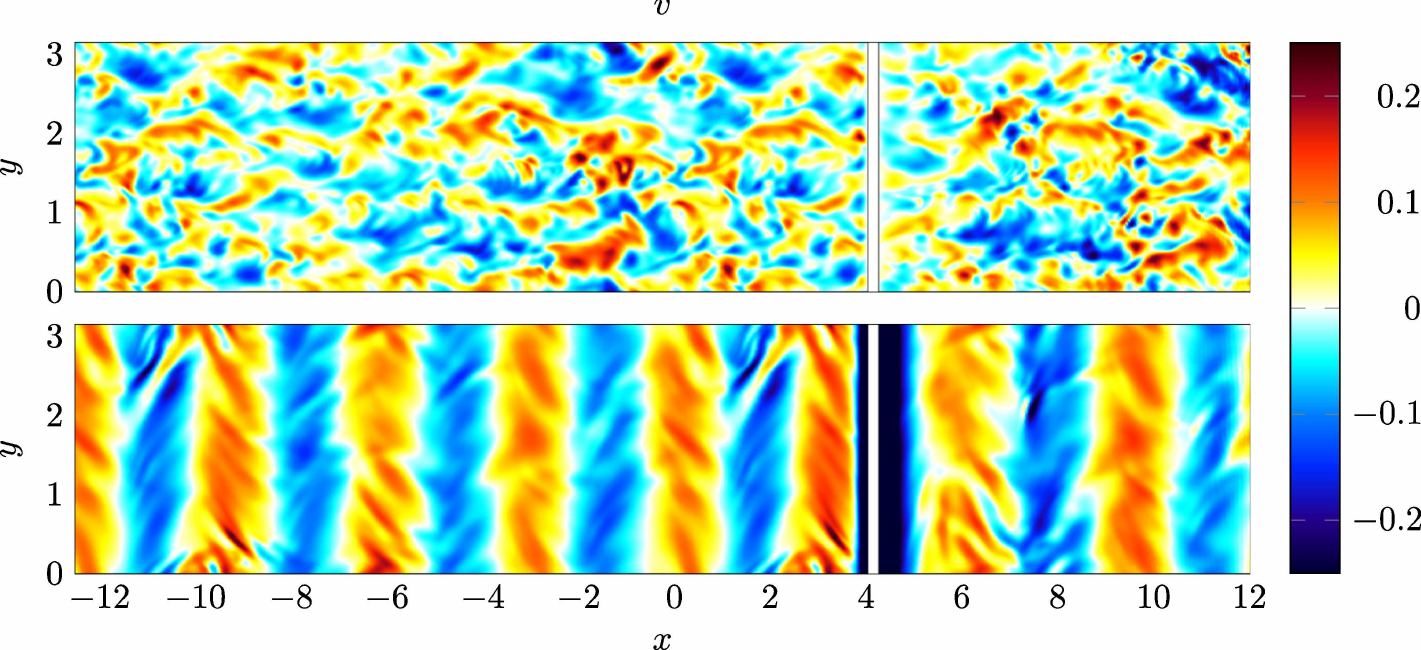}
\caption{Colour plot of an instantaneous spanwise velocity field, in the plane $z = 0.08$ over the bump $G_1$. Panels as in figure \ref{fig:sez-u}.}
\label{fig:sez-v}
\end{figure}

Figures \ref{fig:sez-u} and \ref{fig:sez-v} show instantaneous colour plots of the streamwise and spanwise velocity components in the plane $z=0.08$, which lies just below the bump tip. Every figure compares the flow with (bottom) and without (top) StTW; the white vertical band shows the intersection of the cut plane with the bump. In figure \ref{fig:sez-u}, the elongated streaks of high/low streamwise momentum are clearly visible for the reference simulation upstream of the bump and immediately downstream. In the StTW case, by comparison, the overall velocity level is lower, and the range of fluctuations more limited overall, with a less evident streaky pattern. On the other hand, the streamwise modulation induced by StTW is noticed, particularly just upstream and downstream of the bump tip, where the distance between the wall and the cut plane is small.

In figure \ref{fig:sez-v}, the spanwise velocity over the same $z$ plane is plotted. In the reference simulation the flow organization in turbulent structures can be appreciated, whereas a different picture emerges in the wall-forced case, where the spanwise forcing creates alternating bands of positive and negative spanwise velocity.

\begin{figure}
\centering
\includegraphics[width=\textwidth]{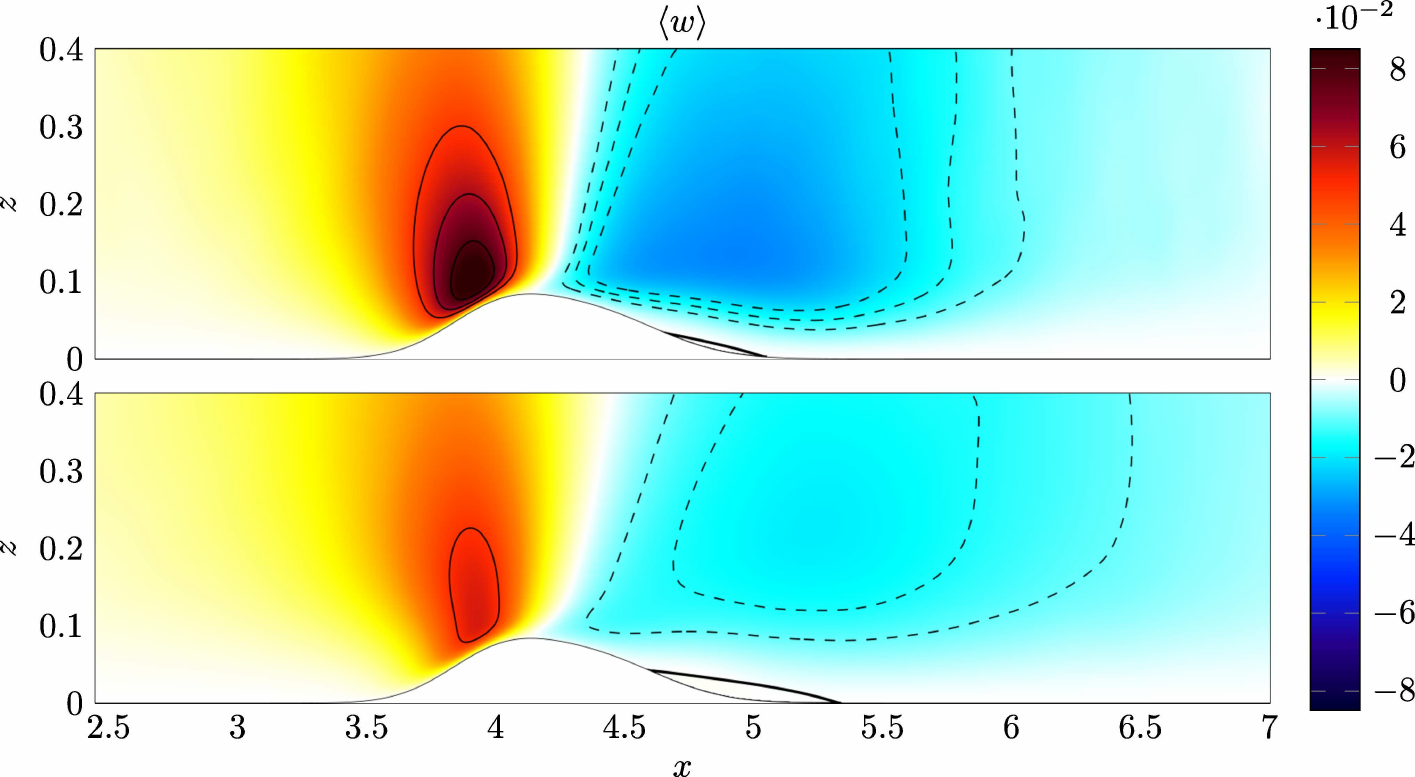}
\caption{Colour plot of the mean vertical velocity $\aver{w}$ for the bump $G_1$: top, reference case; bottom, StTW. Positive contours (continuous lines) are drawn for $\aver{w} = \left(0.05,0.065,0.08\right)$, and negative contours (dashed lines) are drawn for  $\aver{w} = \left(-0.02, -0.015, -0.01\right)$. The thick black line indicates $\aver{u}=0$ and marks the boundary of the separated region.}
\label{fig:w}
\end{figure}

Moving on to the analysis of the mean flow field, for which the operator $\aver{\cdot}$ implies averaging over time and the homogeneous spanwise direction, figure \ref{fig:w} plots a vertical plane with a colour map of the vertical velocity component $\aver{w}$, for a localized portion of the domain which includes the bump, namely $2.5 \le x \le 7$. The plot shows that the peak of $w$ just ahead of the bump is decreased because of StTW. The thick contour line corresponds to $\aver{u}=0$ and visualizes the separated region with a recirculation bubble after the bump; the separated region is very small for the reference case with $G_1$, but the wall forcing somewhat increases its extension.

\begin{figure}
\centering
\includegraphics[width=\textwidth]{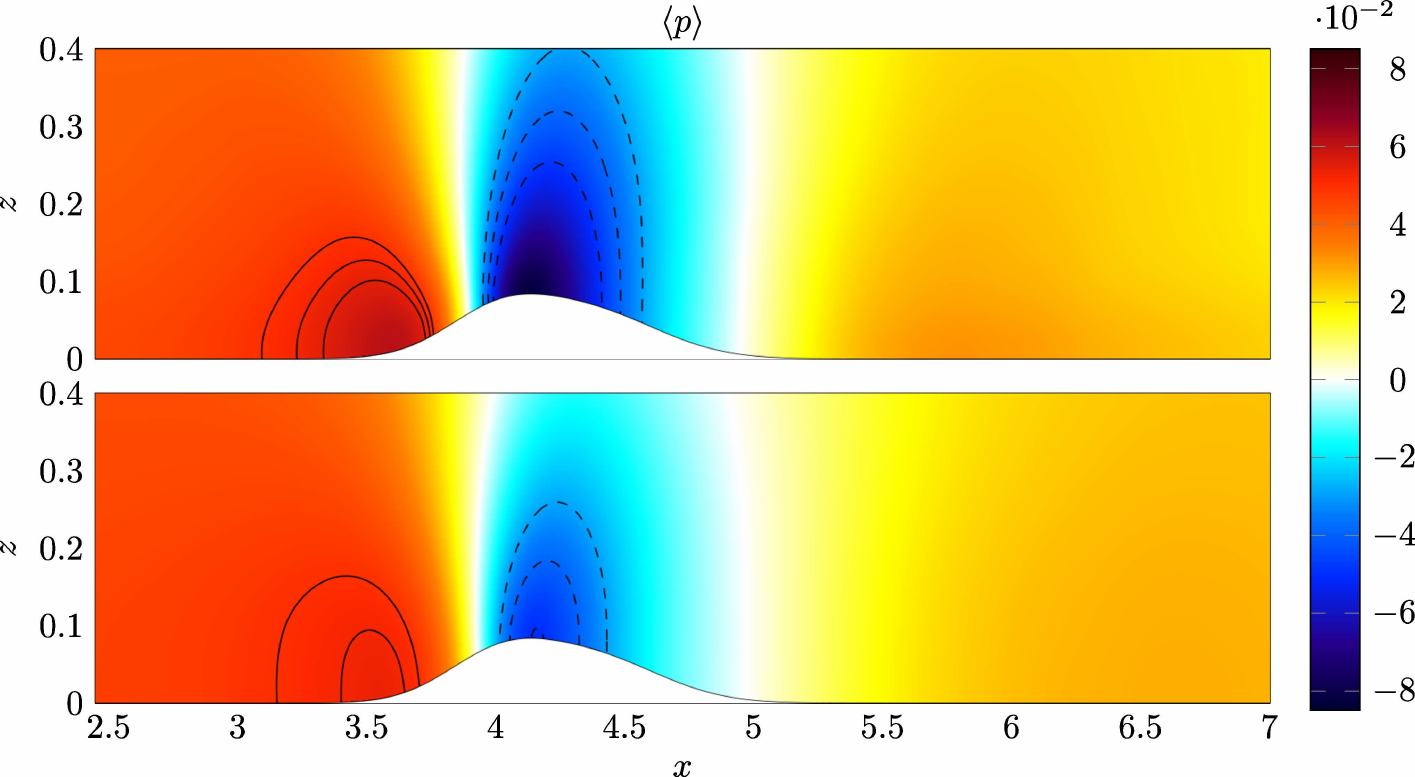}
\caption{Colour map of the mean pressure $\aver{p}$ for the bump $G_1$: top, reference case; bottom, StTW. Positive contours (continuous lines) are drawn for $\aver{p} = \left(0.05,0.0525,0.055 \right)$, and negative contours (dashed lines) are drawn for $\aver{p} = \left(-0.05, -0.04, -0.03\right)$.}
\label{fig:pressure}
\end{figure}

The mean pressure distribution $\aver{p}$ is shown in figure \ref{fig:pressure}; to ease comparison, the pressure levels of the two cases are offset such that they coincide at $x=0$. Pressure locally increases over the anterior part of the bump, while a local minimum appears shortly after the tip because of the negative wall curvature. In the StTW case, the local maximum before the bump is decreased, and similarly the local minimum at the bump tip shows lesser intensity.

%%%%%%%%%%%%%%%%%%%%%%%%%%%%%%%%%%%%%%%%%%%%%%%%
\section{Skin friction and pressure at the wall}
\label{sec:distributions}

The aerodynamic force includes contributions from friction and pressure. Friction and pressure at the wall are often expressed by local dimensionless coefficients, although their integral contribution is not straightforwardly related to the drag force. The friction coefficient is
\begin{equation}
c_f(x) = \frac{2 \aver{\tau_w}(x)}{\rho U_b^2} ,
\label{eq:cf}
\end{equation}
and the pressure coefficient is
\begin{equation}
c_p(x) = \frac{2 \aver{p}(x)}{\rho U_b^2} .
\label{eq:cp}
\end{equation}

In definition \eqref{eq:cf}, $\rho$ is the fluid density and $\tau_w = \mu \ \vect{\hat{t}} \cdot \partial \vect{u} / \partial n$, with $\vect{\hat{t}}$ the tangential unit vector and $ \partial / \partial n$ the derivative in wall-normal direction. In definition \eqref{eq:cp}, the pressure value $p(x)$, which can be arbitrarily shifted in an incompressible flow, is set to have $\aver{p}=0$ at the outlet section.

\begin{figure}
\includegraphics[width=\textwidth]{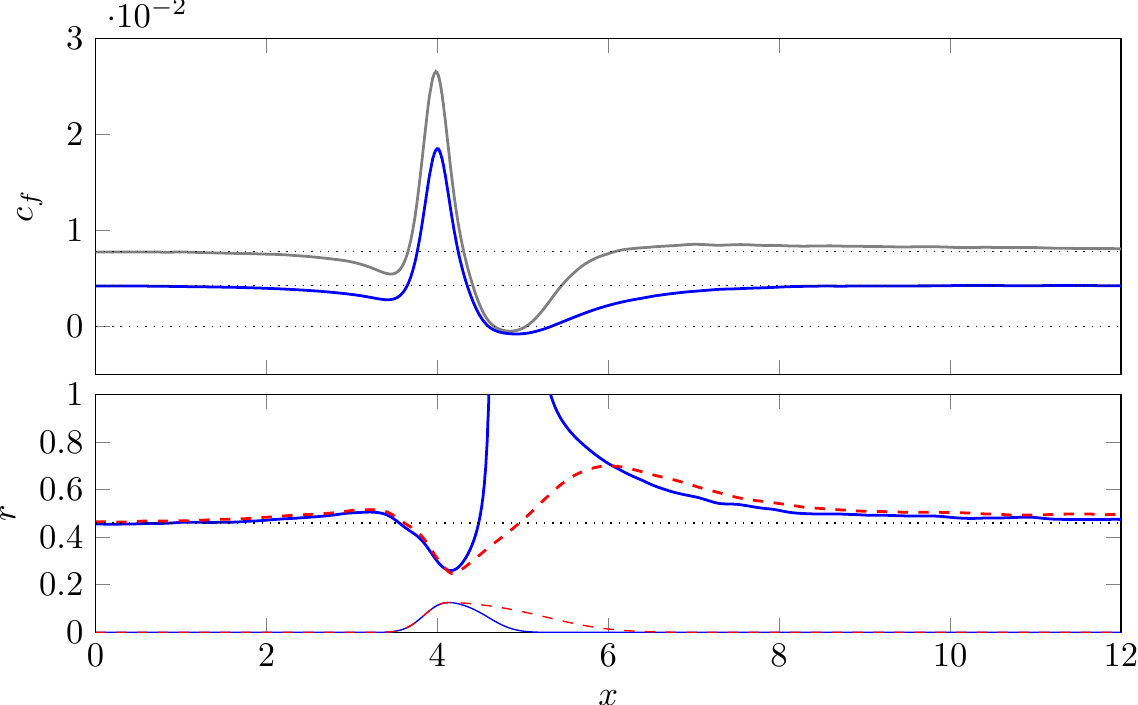}
\caption{Skin-friction distribution $c_f(x)$ over the wall with the bump. Top: comparison between the reference case (grey) and the controlled case (blue) for bump $G_1$. Bottom: local skin-friction reduction rate $r(x)$ for $G_1$ (blue) and $G_2$ (red dashed). The thin profiles at the bottom of the plots draw the two bumps, in arbitrary vertical units.}
\label{fig:Cf-R}
\end{figure}

The quantity $c_f(x)$ is considered in figure \ref{fig:Cf-R} for the reference and controlled cases. Only the lower wall of the non-periodic portion $0 \le x \le 12$ of the computational domain is shown. Indeed, the presence of the bump is felt on the $c_f$ distribution at the opposite wall too; however, owing to the small blockage this effect is minimal and therefore not shown here. The top panel plots the distribution of $c_f(x)$ itself, comparing the reference and the actuated flows for the bump $G_1$. In the reference simulation the friction coefficient decreases just before the bump, and then quickly grows to reach its maximum close to the bump tip. The maximum value is approximately 3 times that of the flat wall. Downstream of the tip, $c_f$ quickly drops towards zero. The flow separation (already discussed in figure \ref{fig:w}), produces a locally negative $c_f$. After reattachment the friction distribution presents a mild overshooting, followed by a slow recovery towards the undisturbed planar-wall value. When StTW are applied, the behaviour of $c_f(x)$ is qualitatively similar, but quantitative changes are introduced, as friction is reduced everywhere by StTW. To quantify such changes, a local skin-friction reduction rate $r(x)$ is plotted in the lower panel for both bumps. $r$ is defined as the relative change of $c_f(x)$ between the controlled and the reference flow:
\begin{equation}
r(x) = 1- \frac{c_{f}(x)}{c_{f,0}(x)}
\end{equation}
where $c_{f,0}(x)$ is for the reference flow. Way upstream of the bump $G_1$, where the wall is flat, $r$ equals the value typical of StTW in the indefinite plane channel flow, namely $r = 46\%$ at the present $Re$ and for the employed parameter values \citep[see for example][]{gatti-quadrio-2016}. When the bump is approached, $r$ at first increases slightly above $50\%$ immediately upstream of the bump, and then decreases to 25\% over the anterior part of the bump. After the tip, when flow separation takes place, the quantity $r$ becomes meaningless: for example, the extrema of the separation bubble in the unforced case, identified by the zero points for $c_{f,0}$, correspond to points where $r$ diverges to infinity. StTW are observed to cause an increase of both intensity and length of the separation bubble. After the reattachment point, differently from the reference case, no overshooting occurs for friction over StTW.

The longer bump $G_2$, though very similar, does not lead to flow separation. Here $r$ (red dashed line) is nearly identical to that for $G_1$ up to the bump tip, and then increases towards a local maximum of $70\%$ near $x=6$. Once again, the recovery towards the planar-wall value is quite slow, and the local drag reduction remains higher than the planar value in most of the computational domain after the bump tip. 

\begin{figure}
\centering
\includegraphics[width=\textwidth]{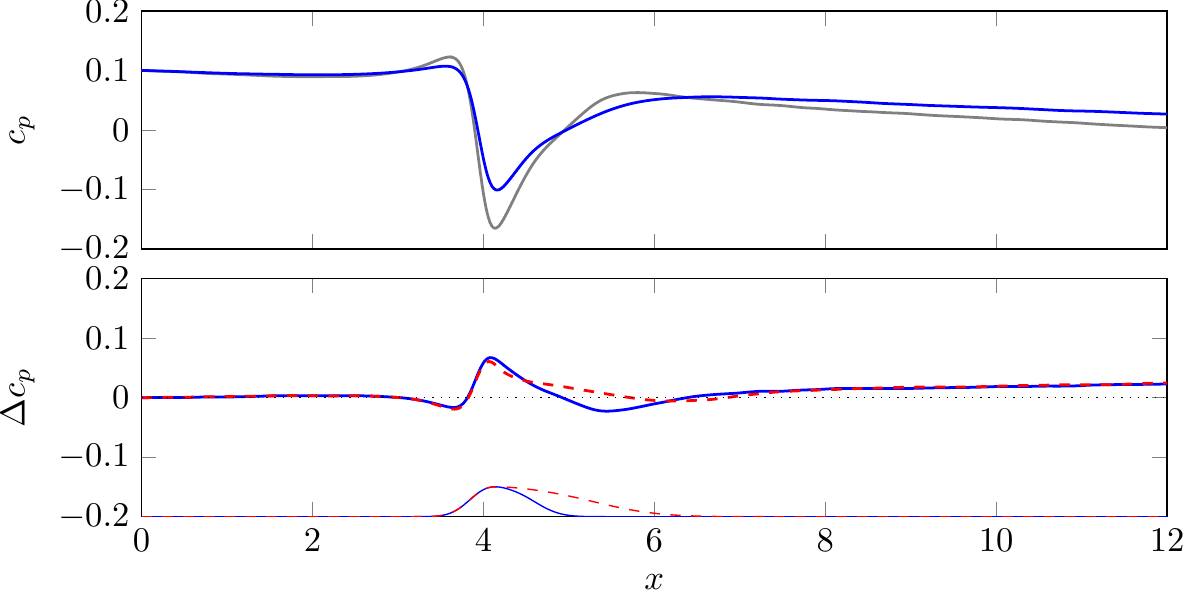}
\caption{Pressure distribution $c_p(x)$ over the wall with the bump. Top: comparison between the reference case (grey) and the controlled case (blue) for bump $G_1$. Bottom: local difference between pressure coefficients $\Delta c_p(x) = c_p(x) - c_{p,0}(x)$ for $G_1$ (blue) and $G_2$ (red dashed). The thin profiles at the bottom of the plots draw the two bumps, in arbitrary vertical units.}
\label{fig:Cp-Delta}
\end{figure}

Figure \ref{fig:Cp-Delta} plots the streamwise distribution of $c_p(x)$; the pressure levels of the two cases, which by definition coincide at the outlet section where the mean pressure is set at zero, have been adjusted such that they coincide at $x=0$ instead, as done already in figure \ref{fig:pressure}. In the reference case, the local pressure increases before the bump, and so does the pressure coefficient. An evident minimum of $c_p$ is reached at the bump tip, followed by a relatively fast recompression. In the inlet and outlet portions of the computational domain, i.e. far enough from the bump, $c_p(x)$ presents the linear decrease (i.e. uniform mean pressure gradient) that is expected for a plane channel flow.  As already commented upon for $c_f$, the bump affects the pressure coefficient on the opposite planar wall too, but this is not shown here as the changes are minimal.

The action of StTW at the inlet, where the local drag is friction-dominated, simply translates into a milder negative slope of the $c_p(x)$ curve, owing to the lower friction drag. More downstream, the positive pressure peak before the bump is noticeably reduced by StTW, thus anticipating that the pressure drag contribution associated to the anterior part of the bump will be reduced (see later \S\ref{sec:budgetp}). The minimum of $c_p$ near the bump tip is also decreased by StTW, so that the pressure jump between the two local extrema is reduced by around $20\%$. After the bump, say for $x>7$, the flow is not affected by the presence of the bump, and again only a reduced pressure gradient is visible.

In order to quantify changes in pressure distribution due to StTW, a local pressure reduction rate, in analogy with friction, cannot be used: in incompressible flows, pressure values can be shifted by a constant, making a ratio meaningless. For this reason we simply study the difference introduced by StTW as
\begin{equation}
\Delta c_p(x) = c_p(x) - c_{p,0}(x).
\label{eq:dcp}
\end{equation}
Figure \ref{fig:Cp-Delta} (bottom panel) plots $\Delta c_p(x)$ for both geometries showing a very similar behaviour; the two curves almost coincide up to the bump tip, then for $G_2$ the milder slope in the aft part creates a slower recompression. The agreement of $\Delta c_p(x)$ up to the bump tip for the two geometries indicates that the effect of StTW on the pressure distribution is not dictated by the presence of a separation bubble.

%%%%%%%%%%%%%%%%%%%%%%%%%%%%%
\subsection{Drag coefficients}
\label{sec:cd}

To assess how StTW interact with the curved wall, simply comparing the drag force per unit area between the plane geometry and the one with the bump is not the best choice, because the bump introduces concentrated losses. In fact, in the limit of very large streamwise extent $L_x^{np}$ of the computational domain that includes the bump, concentrated losses become negligible and the same drag of the planar case is obtained.

Drag is usually quantified in two ways. The first one, meant to evaluate distributed losses $C_{d}^d $, uses the streamwise length of the domain under consideration as the reference length, and leads to the definitions below for friction and pressure contributions:
\begin{equation}\label{eq:cdc}
C_{d,f}^d = \frac{2}{\rho U_b^2 \ L_x} \ \vect{\hat{x}} \cdot \int_0^{L_x} \mu \left( \nabla \vect{u} + \nabla \vect{u}^T \right) \cdot \vect{n} \ \ud \ell ; \qquad C_{d,p}^d = \frac{2}{\rho U_b^2 \ L_x} \ \vect{\hat{x}} \cdot \int_0^{L_x} \aver{p} \vect{n}  \ \ud \ell,
\end{equation}
where $\vect{\hat{x}}$ is the unit vector in the $x$ direction. 
Obviously, in the flat channel only the friction contribution is present. (In the above definitions, note the use of capital letters to indicate global force coefficients, whereas lowercase was used above for local coefficients.)

To evaluate concentrated losses $C_{d}^c $, on the other hand, the obstacle contribution to drag is singled out computing the drag variation with respect to the planar case. Using the frontal area of the obstacle as a reference surface, or in this case the bump height $h_b$ as the reference length, $C_{d}^c $ becomes independent from the domain length:
\begin{equation}\label{eq:cdb}
 C_{d,f}^c = \frac{L_x^{np}}{h_b} \left( \widetilde{C}_{d,f}^d - \overline{C}_{d,f} ^d \right) ; \qquad
 C_{d,p}^c = \frac{L_x^{np}}{h_b} \left( \widetilde{C}_{d,p}^d - \overline{C}_{d,p}^d \right) ,
\end{equation}
where $\widetilde{C}_{d}^d$ and $\overline{C}_{d}^d$ indicate drag coefficients computed for the non-planar and planar wall respectively. Obviously, when concentrated losses $C_{d}^c$ are evaluated for the controlled case, both $\widetilde{C}_{d}^d$ and $\overline{C}_{d}^d$ are computed in presence of StTW.

\begin{table}
\centering 
\begin{tabular}{c |  c c  c | c  c c   }
                 &   \multicolumn{3}{c|}{Distributed losses}   & \multicolumn{3}{c}{Concentrated losses}  \\
                             & Ref           &  StTW    & $\Delta$  &  Ref   & StTW   &  $\Delta$  \\
\midrule 
$C_{d,f} \times 10^{-2}$     & $0.777$       & $0.424$  & $-45.5\%$  & $-0.004$ & $-4.671$ & $ $ \\         
$C_{d,p} \times 10^{-2}$     & $0$           & $0$      & $0$       &  $9.891$  & $8.887$ & $-10.3\% $  \\        
$C_d   \times 10^{-2}$       & $0.777$       & $0.424$  & $-45.5\%$ &  $9.887$  & $ 4.197$ & $-57.5\% $    \\
\end{tabular}
\caption{Drag coefficients for the bump $G_1$. $C_{d,f}$  and $C_{d,p}$ are the friction and pressure components respectively, with $C_d=C_{d,f}+C_{d,p}$. Distributed losses are computed according to Eq. \eqref{eq:cdc} in the planar geometry while concentrated losses introduced by the bump are evaluated via Eq.\eqref{eq:cdb}. Figures are for the lower wall only.}
\label{tab:Cd1}
\end{table}

In table \ref{tab:Cd1}, drag coefficients are reported for the bump $G_1$, computed over the lower wall only. Distributed losses are evaluated in the flat periodic domain while concentrated losses are computed in the non-periodic domain that includes the bump. In the planar case StTW reduce friction drag by $46\%$, as expected, and no pressure drag is present. The friction component of the concentrated losses due to the bump is globally nearly zero in the reference case, implying that the friction coefficient computed over the entire wall with the bump almost coincide with that over the flat wall. This is non obvious, as the bump has been observed (cfr. Figure \ref{fig:Cf-R}) to introduce significant local variations. The pressure component, on the other hand, generates a considerable additional contribution to drag. 

When StTW are applied, concentrated friction losses become negative, implying that the mean friction over the wall with the bump is lower than the friction over a controlled plane wall. This benefit, absent without flow control, is due to the slower downstream recovery of friction to its planar value and the lack of overshooting after the bump, as shown in figure \ref{fig:Cf-R}. However, quantifying this benefit in terms of the percentage change of $C_{d,f}^c$ (not shown in Table \ref{tab:Cd1}) would be meaningless, since the reference value is close to zero. The bump is responsible for a considerable pressure drag, and it is interesting to observe that StTW reduce this component too, by an amount of approximately $10\%$. Overall, control by StTW leads to reduction of total concentrated losses by around $57\%$. 

\begin{table}
\centering 
\begin{tabular}{c |  c c  c | c  c c   }
                 &   \multicolumn{3}{c|}{Distributed losses} &  \multicolumn{3}{c}{Concentrated losses}  \\
                       & Ref           &  StTW    & $\Delta$        & Ref   & StTW   &  $\Delta$ \\   
\midrule 
$C_{d,f} \times 10^{-2}$     & $0.781$       & $0.418$  & $-46.5\%$   &  $-0.158$ & $-2.904$ & $ $ \\         
$C_{d,p} \times 10^{-2}$     & $0$           & $0    $  & $ 0     $   &  $7.083$  & $ 6.843$ & $-3.4\% $  \\           
$C_d   \times 10^{-2}$     & $0.781$       & $0.418$  & $-46.5\%$   &  $6.925$  & $ 3.940$ & $-43.1\% $    \\
\end{tabular}
\caption{Drag coefficients for the bump $G_2$.}
\label{tab:Cd2}
\end{table}

Table \ref{tab:Cd2} provides the same quantities for the milder bump $G_2$. The distributed losses in the planar case are obviously the same as $G_1$ (the small difference is attributed to the finite averaging time). Since $G_2$ is less steep after the tip, the concentrated losses decrease. The concentrated friction losses are clearly negative; pressure recovery is more effective and thus pressure losses are lower in both reference and controlled case. The presence of StTW also induces a pressure drag reduction, which is however less pronounced than in $G_1$. The overall outcome is a reduction of about $43\%$ in the concentrated losses. 

It is noted explicitly that the reduction of concentrated losses reported in the rightmost column of tables \ref{tab:Cd1} and \ref{tab:Cd2} must be added to the distributed drag reduction to assess the global saving. This is discussed in the following paragraph.

%%%%%%%%%%%%%%%%%%%%%%%%%%%%%%%%%%%%%%%%%%%%%%%%%%%%%%%%%%%%%%%%%%%%%%
\subsection{Changes in friction and pressure drag over the curved wall}

Drag changes induced by StTW are now assessed against the scenario in which StTW are assumed to simply reduce the friction component by the amount they would in a plane channel. We name this the "extrapolated amount", and indicate it with an $(e)$ superscript. 

The simulation for $G_1$ shows that the friction drag over the entire non-periodic portion is 92\% of the overall drag. Combining changes in distributed and concentrated losses, StTW reduce friction drag by $49.6\%$, i.e. approximately 4\% more than the planar case. The pressure drag, representing 8\% of the total drag, is reduced by a relative 10.3\%. Bump $G_2$, albeit separation-free, shows a similar behavior. With friction drag accounting for 94\% of the total, StTW reduce friction component by 3\% more than the planar value, and the pressure component too is reduced by a further 3\%.

A physical explanation of the inter-relation between friction and pressure drag reduction requires quantifying the influence of local stresses on the global drag budget. A local drag reduction coefficient $\Delta c_d(x)$ is defined as the difference of local contributions to drag coefficients, i.e. the integrands in Eqns.\eqref{eq:cdc}, between the reference and the controlled case:
\begin{equation}
\Delta c_{d}(x)  =  c_{d,0}(x) - c_d(x) ,
\label{eq:deltacd}
\end{equation}
where the subscript $0$ indicates the reference case. It should be noted that a sign inversion is adopted here in comparison to Eq. \eqref{eq:dcp}, for  drag benefits to yield positive $\Delta c_{d}(x)$. To weigh the local contribution on the integral budget, a global drag reduction rate $R(x)$ is further introduced as in \cite{stroh-etal-2016}: $R(x)$ is defined as the integral of $\Delta c_{d}(x)$ up to location $x$, expressed as a percentage of the drag calculated over the entire domain in the reference case:
\begin{equation}
R(x)  =  
\frac{\int_0^x \Delta c_d(x') \ud x' }{\int_0^{L_x}  c_{d,0}(x') \ud x' } .
\label{eq:integratedR}
\end{equation}

%-----------------------------------
\subsubsection{Friction drag reduction}

\begin{figure}
\centering
\includegraphics[width=\textwidth]{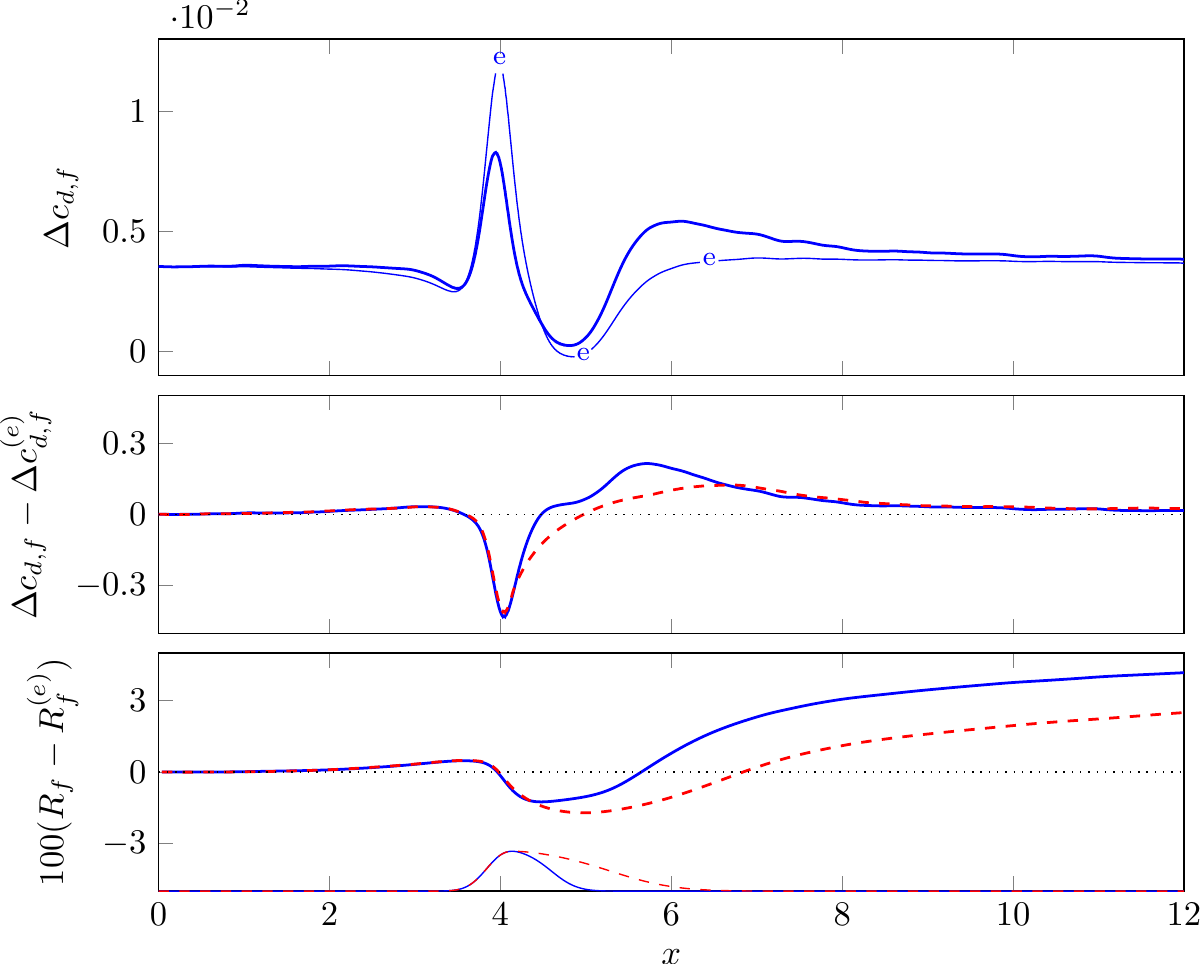}
\caption{Changes in the skin-friction component of the total drag. Top: the computed $\Delta c_{d,f}$ (thick line) compared with the extrapolation $\Delta c_{d,f}^{(e)}$ (thin line with labels) for bump $G_1$. Center: difference between computed and extrapolated friction drag reduction, for geometries $G_1$ (blue line) and $G_2$ (red dashed line). Bottom: difference between actual  $R_f$ and extrapolated integral budget $R_f^{(e)}$ for both geometries. The thin profiles at the bottom of the plot draw the two bumps, in arbitrary vertical units.}
\label{fig:Dcdf}
\end{figure}

The friction component of $\Delta c_d$, namely $\Delta c_{d,f}$, is plotted in the upper panel of figure \ref{fig:Dcdf}, together with the extrapolated value $\Delta c_{d,f}^{(e)}$ obtained by assuming that the planar friction reduction rate carries over to the non-planar geometry $G_1$. Obviously, the two curves tend to coincide far from the bump, while immediately upstream and over a large downstream extent the true friction drag reduction is larger than the extrapolated one. On the other hand, over the anterior part of the bump, where $r$ in figure \ref{fig:Cf-R} shows a local minimum, the true friction drag reduction is smaller than $\Delta c_{d,f}^{(e)}$.

To understand which area is specifically responsible for the extra friction drag reduction, the local difference between actual and extrapolated values is shown in the center panel of figure \ref{fig:Dcdf}, for both bump geometries, to demonstrate that the qualitative behaviour is the same, regardless of the presence of flow separation. The two curves coincide, as expected, at the inlet and in the fore part of the bump, whereas the lack of separation in $G_2$ makes the two curves quantitatively differ in the decelerating region after the bump tip. The integral budget $R_f$, i.e. the friction component of the integrated drag reduction introduced by Eq.(\ref{eq:integratedR}), is plotted in the lower panel of figure \ref{fig:Dcdf} as a difference with respect to the extrapolated value $R_f^{(e)}$. For both bumps, the larger friction drag reduction after the bump tip, and its slower recovery of the planar value, already discussed in the context of figure \ref{fig:Cf-R}, translate into a global friction drag reduction approximately 3\% larger than the value extrapolated from the planar case.

\begin{figure}
\centering
\includegraphics[width=\textwidth]{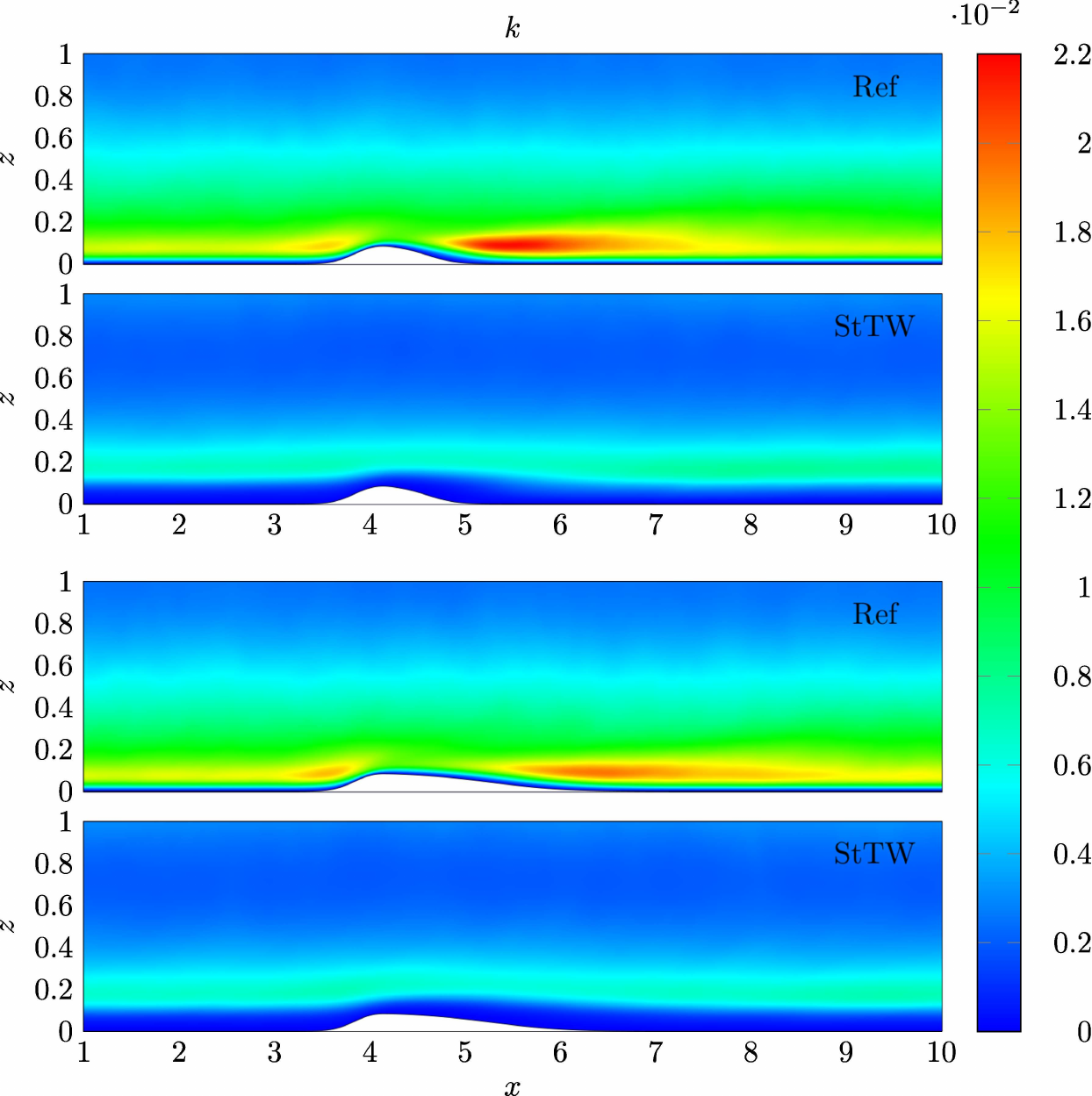}
\caption{Colour plot of the turbulent kinetic energy, in outer units, for the bump $G_1$ (top) and $G_2$ (bottom), with and without StTW.}
\label{fig:k}
\end{figure}

\begin{figure}
\centering
\includegraphics[width=\textwidth]{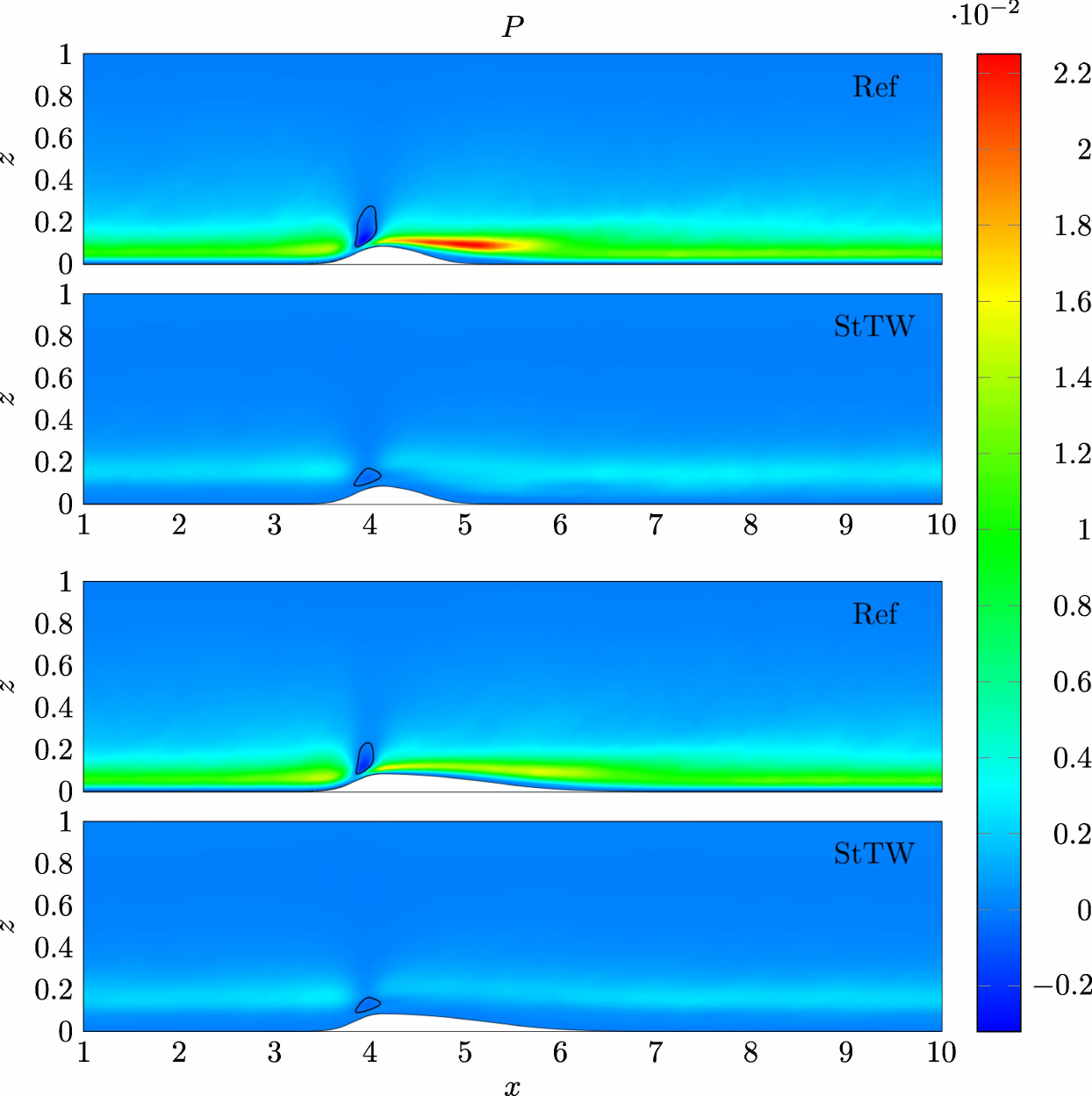}
\caption{Colour plot of the production $P$ of turbulent kinetic energy. The level $P=0$ is indicated by the contour line. Panels as in figure \ref{fig:k}. }
\label{fig:Prod}
\end{figure}

Figure \ref{fig:k} links the differences in friction drag reduction observed after the bump tip, including where the wall is flat again, to the distribution of the turbulent kinetic energy $k =  1/2 \aver{u_i' u_i'}$. Its production $P$ is shown later in figure \ref{fig:Prod}. It is worth mentioning that the fluctuating velocity field $u_i'$ is defined by subtracting the local mean field and, for the StTW case, by employing a phase average to additionally remove the contribution of the spanwise Stokes layer. 

In agreement with literature, for the reference cases figure \ref{fig:k} shows two areas of high $k$: one just ahead of the bump, and the other, more intense, immediately after the bump tip, extending approximately one bump length and related to the strong adverse pressure gradient \citep{wu-squires-1998}. \cite{marquillie-ehrenstein-laval-2011} discuss a similar picture based on the streamwise distribution of the maximum of turbulent kinetic energy. For the milder bump $G_2$ the strong peak after the tip is weakened, whereas the local maximum before the tip is unchanged.  The controlled cases show both quantitative and qualitative differences. At the inlet, in agreement with the observations by \cite{quadrio-ricco-2011} for the flat wall, the maximum value of $k$ (measured in outer units) is reduced by StTW and displaced at larger wall distances, and small values of $k$ are observed within the Stokes layer. The interaction with the bump appears to be minimal, with the peak value of $k$ remaining nearly constant along the streamwise coordinate, suggesting that the Stokes layer effectively hinders the propagation of the geometrical perturbation made by the bump into the buffer layer and above. There appears to be no substantial difference between $G_1$ and $G_2$.

In figure \ref{fig:Prod} the production $P$ of turbulent kinetic energy: 
\begin{equation}
P =  - \aver{u_i' u_j'}\frac{\partial \aver{u_i}}{\partial x_j}
\end{equation} 
is plotted to show that the spatial distribution of $k$ is consistent with that of its production: even for $P$, the strong streamwise variation of the reference case is significantly altered by the StTW. Over the bump without actuation, as already observed by \cite{mollicone-etal-2017}, the turbulent production increases slightly before the bump and then drops to slightly negative values in the accelerated region just before the bump tip. A large positive peak of turbulent production follows, beginning at the bump tip. The intensity of the local maxima is lower for $G_2$. When StTW are applied, the two cases show a similar behavior: streamwise changes of $P$ are strongly inhibited, and only a local slightly negative minimum in the accelerating region can be detected, though both the extension and the absolute value of the minimum are considerably reduced. In addition, careful scrutiny of the various contributions to $P$ (not shown) reveals that the major cause for the difference between the uncontrolled and controlled flows rests with the field of the Reynolds stresses, while the gradients of the mean flow are much less affected.  

The increase of turbulent activity is related to the friction increase producing the overshoot in the $c_f$ curves of figure \ref{fig:Cf-R} downstream of the bump. Figures \ref{fig:k} and \ref{fig:Prod} confirm that, with StTW, turbulent activity is inhibited and no overshooting is found in the friction distribution. It is the area immediately downstream the bump tip (see figure \ref{fig:Dcdf}) that is associated with the extra friction reduction.

%----------------------------------------------------------------------------------------------------------------------------
\subsubsection{Pressure drag reduction}
\label{sec:budgetp}

\begin{figure}
\centering
\includegraphics[width=\textwidth]{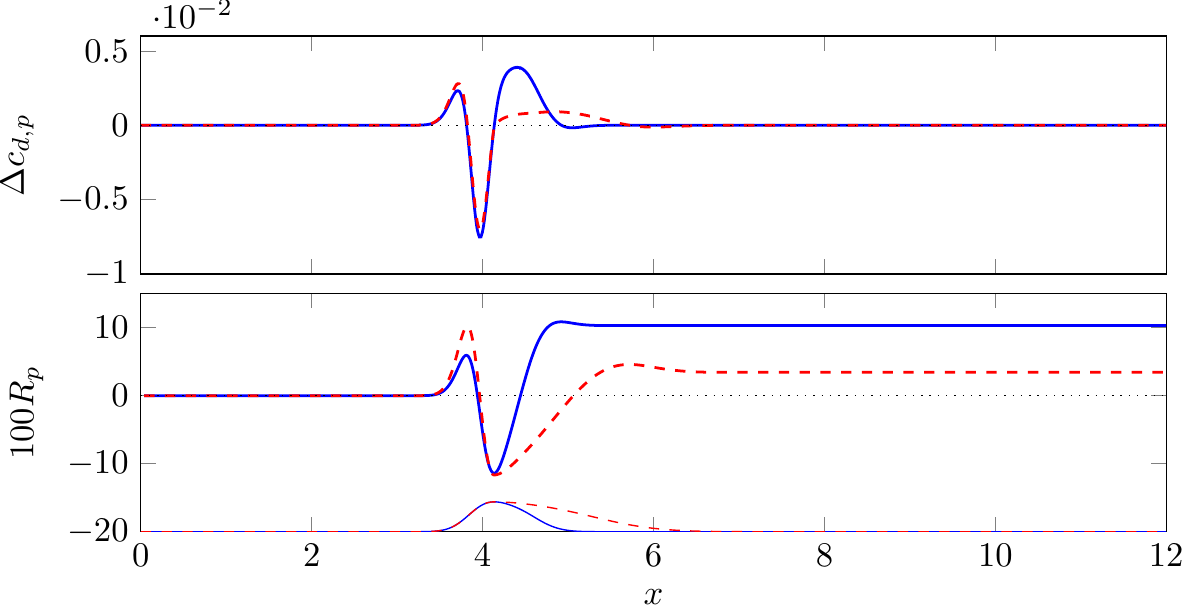}
\caption{Comparison of the contribution to pressure drag  changes between $G_1$ (blue) and $G_2$ (red dashed). Top:  $\Delta c_{d,p}(x) $; bottom: integral budget $R_p$ for both geometries. The thin profiles at the bottom of the plot draw the two bump geometries.}
\label{fig:Dcdp}
\end{figure}

A similar analysis is now carried out for pressure drag, for which simple extrapolation from the planar case would indicate no reduction at all. Figure \ref{fig:Dcdp} examines first the streamwise distribution of $\Delta c_{d,p}(x)$, the pressure component of (\ref{eq:deltacd}). The upper panel plots $\Delta c_{d,p}(x)$ for both bumps, and the lower panel shows $R_p(x)$, the pressure component of (\ref{eq:integratedR}). 

As already shown in figure \ref{fig:Cp-Delta}, StTW reduce the positive pressure peak upstream of the bump, as well as the negative one near the bump tip. In terms of drag, the reduction of the first peak is beneficial, and translates into a local drag reduction and a positive $\Delta c_{d,p}(x)$ with a similar (albeit not identical) local maximum for both geometries, located at the same streamwise coordinate. Such agreement between the two geometries implies that the reduction of pressure drag in the anterior part of the bump is not related to the changes in the separation bubble. The attenuation of the first pressure peak alone produces 6\% of pressure drag reduction for the geometry $G_1$ and 10\% for $G_2$. The reduction in the intensity of the second, negative pressure peak starts upstream of the bump tip, and extends downstream. Because of the orientation of the surface normal, before the tip such changes are detrimental to drag reduction, and  become beneficial after the tip. The negative peak of $\Delta c_{d,p}(x)$ is essentially identical for both geometries up to the bump tip; however, in the region downstream of the tip, the different local slope of the wall implies a different projection of wall-normal force in the horizontal direction. For this reason, the milder bump $G_2$ only partially benefits from the increased pressure recovery in the aft part of the bump. The global effect is therefore a 10\% pressure drag reduction for $G_1$, and only $3\%$ for $G_2$, as already shown in Tables \ref{tab:Cd1} and \ref{tab:Cd2}.

Pressure drag reduction has been shown to be unrelated to changes in the separation bubble produced by $G_1$. However, it is interesting to explore such changes, although they produce no significant effect in the global drag budget. The streamwise extent of the separated region is determined by looking at the zeroes of the skin-friction distribution shown in figure \ref{fig:Cf-R}. The detached region starts at $x_{d,0}=4.67$ for the reference case and at $x_d=4.6$ for the one with StTW: the two detachment points are very close to each other, with the controlled one moved slightly upstream by the control. The two reattachment points are at $x_{r,0}=5.03$ and $x_r=5.32$ respectively. Overall, StTW produce a longer separated region, with length $L_s=0.72$, twice the length of the reference case $L_{s,0}=0.36$. Equivalent information was already available from figure \ref{fig:w}, where the spatial extent of the recirculating region was determined from the zero of the streamwise component of the velocity. The intensity of the recirculating flow, in terms of largest negative wall shear stress, increases by 60\% with StTW.

\begin{figure}
\centering
\includegraphics[width=1\textwidth]{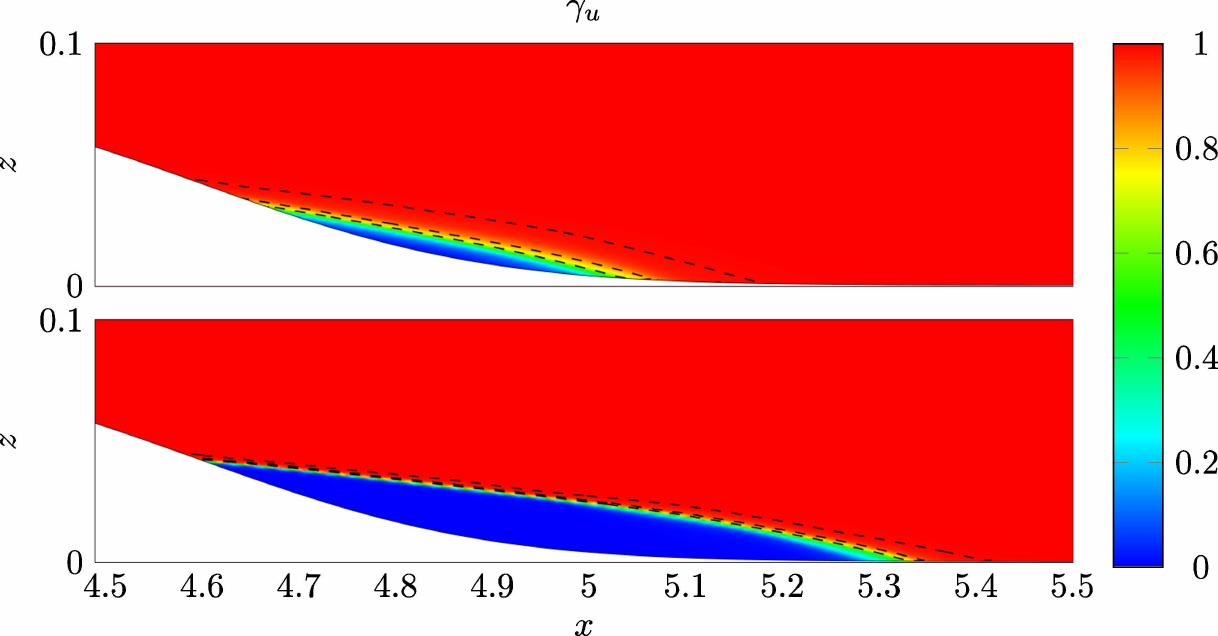}
\caption{Contour of the probability $\gamma_u$ of non-reverse flow, without (top) and with (bottom) StTW over the bump $G_1$. Black lines identify contours of $\gamma_u=(0.50, 0.80, 0.99)$ involved in the definitions by \cite{simpson-chew-shivaprasad-1981a}.}
\label{fig:gamma}
\end{figure}

The inner structure of the separation bubble is further investigated, following \cite{simpson-chew-shivaprasad-1981a}, by locally computing the probability $\gamma_u$ that $u>0$. Incipient detachment is conventionally associated to $\gamma_u=0.99$, i.e. backflow is observed only for 1\% of the time; intermittent transitory detachment takes place when $\gamma_u = 0.80$, transitory detachment when $\gamma_u = 0.50$, and full detachment when $\aver{\tau_w}=0$. Figure \ref{fig:gamma} is a colour plot of $\gamma_u$, along with the contour lines corresponding to the  three values of $\gamma_u$ mentioned above. Moreover, table \ref{tab:gamma} contains quantitative information regarding the detachment and reattachment points, as well as the spatial extent of the separated region.

\begin{table}
\centering 
\begin{tabular}{c | cccccc }
  & $x_{d,0}$  & $x_d$ & $x_{r,0}$ & $x_r$  & $L_{b,0}$ & $L_b$ \\
\midrule 
$\aver{\tau_w} =0$        &  4.67 & 4.60 & 5.03 & 5.32 & 0.36 & 0.72 \\
$\gamma_u = 0.5$  &  4.65 & 4.59 & 5.04 & 5.33 & 0.39 & 0.74 \\
$\gamma_u = 0.80$ &  4.64 & 4.59 & 5.06 & 5.34 & 0.42 & 0.75 \\
$\gamma_u = 0.99$ &  4.58 & 4.58 & 5.18 & 5.40 & 0.60 & 0.82 \\
\end{tabular}
\caption{Detachment and reattachment points for the reference and controlled cases, along with longitudinal extent deduced for specified values of the probability function $\gamma_u$.}
\label{tab:gamma}
\end{table}

The plot confirms that a longer separation bubble is created by StTW, but adds the information that the recirculating region also undergoes qualitative changes. The reference flow presents a diffused interface between the core of the recirculation where $\gamma_u$ is closed to 0 and the attached flow where $\gamma_u=1$, whereas the case with control shows a sharper interface, hinting at a separation bubble that is almost steady and does not undergo temporal oscillations.

%--------------------------------
\section{Global power budget}
\label{sec:budget}

StTW are an active flow control technique, which requires actuation power and is capable to favorably alter the power budget in a turbulent plane channel flow \citep{quadrio-ricco-viotti-2009}. The power budget related to the lower wall is now computed separately for the plane wall (i.e. the periodic simulation that feeds the portion of the channel with the bump) and the bumped wall, and reported in Table \ref{tab:power-G1} for the bump $G_1$. Figures are normalized with the power $P_{tot}$ due to the total drag of the non-actuated case.

\begin{table}
\centering 
\begin{tabular}{c|  c c  c | c  c c  c  }
                 &   \multicolumn{3}{c|}{Plane}    & \multicolumn{4}{c}{Bump}  \\  
                 & Ref &  StTW  & $\Delta$  & Ref   & StTW   &  $\Delta$ & Extrapolated \\ \midrule % \thinhline 
%$P_f$            & $1$   & $0.545$  & $-45.5\%$   & $0.918$ & $0.462$ & $-49.6\% $   & $-45.5\% $\\         
%$P_p$            & $0$   & $0    $  & $ 0     $   & $0.082$ & $0.073$ & $-10.3\% $   & $  0     $\\         
$P_{tot}$        & $1$   & $0.545$  & $-0.455 $   & $1    $ & $0.535$ & $-0.465  $   & $-0.418  $\\        
$P_{req}$        & $-$   & $0.340$  & $+0.340 $   & $ -   $ & $0.312$ & $ +0.312 $   & $+0.313  $ \\        
$P_{net}$        & $-$   & $  -   $  & $-0.115 $   & $ -   $ & $  -  $ & $- 0.153 $   & $-0.105  $  \\        
\end{tabular}
\caption{Power budget for the bump $G_1$. $P_{tot}$ is the power required to overcome the total drag produced by the lower wall. $P_{req}$ is the power required for actuation, and $P_{net}=P_{tot}-P_{req}$ represents the net balance. Figures are for the lower wall only.}
\label{tab:power-G1}
\end{table}

In the plane geometry, where no pressure drag is present, results agree with those by \cite{gatti-quadrio-2016}.  The percentage reduction of pumping power is by definition identical to the change $\Delta C_{d,f}$ of the distributed losses already reported in table \ref{tab:Cd1}. The net power saving is positive but amounts to only $11.5\%$ of the reference total power, since in the present study StTW are made to work to maximize their drag-reducing effect, hence the required actuation power $P_{req}$ is quite large, namely $34\%$ of the total power.

Over the curved wall, StTW reduce friction drag by an amount that exceeds that of the plane wall, and decrease pressure drag too. To quantify their extra benefit, an additional column in table \ref{tab:power-G1} reports the "extrapolated" power budget, obtained via the assumption (already discussed in \S \ref{sec:cd}) that the planar friction reduction carries over to the frictional drag component, with no effect on the pressure component. In table \ref{tab:power-G1}, the total power is reduced by 46.5\% instead of 41.8\%, with a 10\% relative improvement. Since actuation power is almost unchanged, the net savings become 15.3\% instead of 10.5\%, with a relative increment of almost one-half. These specific figures obviously depend on the ratio between friction and pressure drag, i.e. on the bump geometry and the relative extent of the planar surface; a larger difference between actual and extrapolated net power saving can be expected for a larger bump.

\begin{table}
\centering 
\begin{tabular}{c|  c c  c | c  c c  c }
                  &   \multicolumn{3}{c|}{Plane}           & \multicolumn{4}{c}{Bump}    \\     
                  &   Ref        &  StTW            & $\Delta$     & Ref    & StTW   &  $\Delta$ & Extrapolated  \\ \midrule % \thinhline 
%$P_f$             &$ 1 $    & $0.535$  & $ -46.5\% $     & $ 0.940$  & $0.480 $ & $-49.0\% $  & $-46.5\% $\\
%$P_p$             &$ 0 $    & $0    $  & $ 0       $     & $ 0.060$  & $0.058 $ & $-3.4\%  $  & $0      $ \\ 
$P_{tot}$         &$ 1 $    & $0.535$  & $ -0.465  $     & $ 1    $  & $0.538 $ & $-0.463  $  & $-0.437 $ \\
$P_{req}$         &$ - $    & $0.336$  & $ +0.336  $     & $ -    $  & $ 0.312$ & $ +0.312 $  & $+0.317 $ \\ 
$P_{net}$         &$ - $    & $ -   $  & $ -0.129  $     & $ -    $  & $  -   $ & $-0.151  $  & $-0.120 $ \\  

\end{tabular}
\caption{Power budget for $G_2$, as in table \ref{tab:power-G1}.}
\label{tab:power-G2}
\end{table}

The power budget for the bump $G_2$ is reported in Table \ref{tab:power-G2}. The plane wall obviously shows the same figures as the previous case, the minor differences being due to finite averaging time. The non-planar wall is qualitatively similar too, with a total drag reduction of 46.3\%  instead of 43.7\%. The net power saving is 15.1\% of the total power, i.e. larger than the extrapolated value of 12\%. Hence, for both attached and separated flow, StTW produce extra benefits when applied over curved walls.

\begin{figure}
\centering
\includegraphics[width=0.8\textwidth]{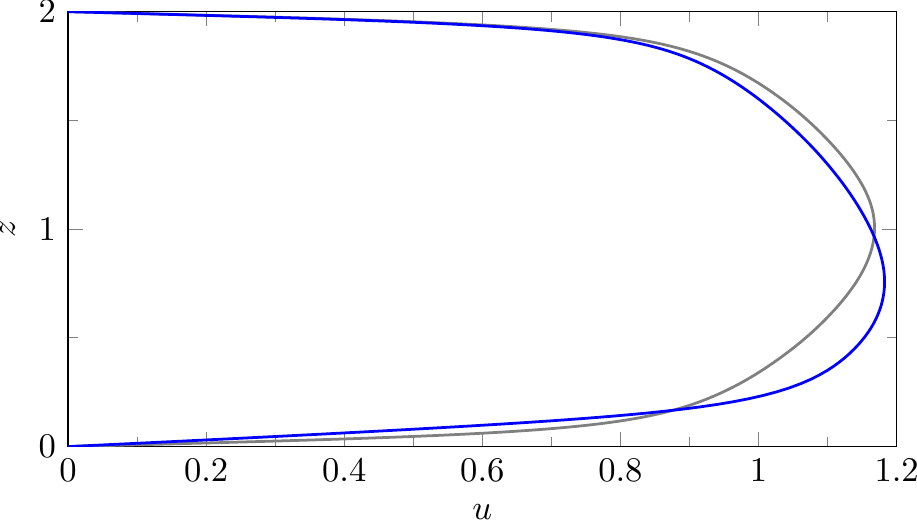}
\caption{Streamwise mean velocity profile in the periodic plane channel. Comparison between reference flow (grey) and actuated flow on the lower wall (blue).}
\label{fig:umean}
\end{figure}

A final comment is about the power budget discussed above, which concerns only the lower wall with the bump. Indeed, additional benefits appear once the upper, flat wall is included. Even though the upper wall has no actuation, StTW applied on the lower wall induce an asymmetry in the mean streamwise velocity profile, so that friction is reduced by 4\% on the upper wall too. This asymmetry, which is also present in the planar case, is explained by the displacement of momentum towards the wall with lower friction. Figure \ref{fig:umean} plots the two mean streamwise velocity profiles, with and without StTW, for the planar case: the velocity maximum is displaced towards the lower actuated wall. This extra benefit on the upper wall, obtained with no additional power, brings the global drag reduction for the whole channel containing the bump $G_1$ to $26\%$, $3\%$ more than the extrapolated value, and a net power savings of 10\%, i.e. nearly doubling the extrapolated value.

%----------------------
\section{Conclusions}
\label{sec:conclusions}
Direct Numerical Simulations of an incompressible turbulent channel flow with a curved wall have been carried out to understand how skin-friction drag reduction affects the total drag. One of the channel walls has a small bump that generates a pressure contribution to the total drag. Two bump geometries are considered, to study cases with and without separation. The flow is modified by a spanwise-forcing technique (streamwise-travelling waves of spanwise wall velocity, or StTW) known to reduce friction drag. Parameters of StTW are tuned to yield a large skin-friction drag reduction of 46\% in the plane case. Friction and pressure distributions over the entire domain length are studied to quantify changes to drag and to the global power budget. The study demonstrates that, for both bumps, the actual power saving obtained by StTW is larger than the extrapolated value obtained by carrying the planar friction reduction over to the friction component of the total aerodynamic drag, while assuming no effect on the pressure component.

In the flow without actuation, friction locally increases in the anterior part of the bump; a local minimum is observed just downstream of the bump tip, with negative values in case of flow separation. The friction then re-increases to reach values slightly larger than the planar one, and eventually recovers slowly, so that a long downstream distance is required to attain the planar value again. When StTW are used, their efficiency varies along the streamwise coordinate, and in particular there is no friction overshooting after the bump, so that a wide region exists where the local friction reduction rate is higher than that of the planar case. 

The pressure distribution is modified by changes in friction. StTW induce a considerable reduction of pressure drag, which amounts to more than 10\% for the cases studied. We have established that pressure drag reduction is not directly linked to flow separation, as it is observed with both bump geometries. When present, however, the separation bubble is significantly affected by the StTW. Indeed, the separated region becomes larger, but at the same time strongly stabilized, almost lacking temporal oscillations.

The combined effect of the StTW upon friction and pressure drag generates a considerable improvement of the global energy-saving performance of StTW. In the simulations described in our study, the amount of net power savings is about one-half larger than in the plane channel alone. If the modifications induced by StTW onto the opposite, non-actuated plane wall are accounted for, the net power savings are increased by 100\%. At any rate, such significant improvements still are underestimates, since these figures strongly depend upon the bump geometry, which has been chosen here without prior knowledge. 

It is not immediate to generalize these results to different geometries, or to different drag reduction techniques. Based on preliminary studies, we can at least vouch for the general picture to remain unchanged when variants of spanwise forcing are employed. However, the main point made by the present work is simply to establish, albeit in a specific case, a concept that sometimes tends to be overlooked: altering the frictional component of the aerodynamic drag in a complex configuration leads to changes in the pressure drag too. This confirms the fundamental idea of recent works \citep[see e.g.][]{mele-tognaccini-catalano-2016} where a RANS-based estimate of the reduction in the overall drag of a modern commercial aircraft covered by riblets was made. Such an estimate has a limited reliability (because of the RANS approach, and because riblets were accounted for indirectly via a modification of the turbulence model at the wall). However, the results seem to indicate that skin-friction drag reduction applied to a body of complex shape provides extra benefits compared to the plane case. This is in agreement with the present DNS-based study, and motivates further research efforts in this direction. In particular, it is intriguing to notice how \cite{mele-tognaccini-catalano-2016} found that the largest beneficial indirect effect from riblets descend from the interaction between the modified skin-friction and the shock wave on the airplane wing. This effect is obviously absent in the present, incompressible flow; work to extend the study to the compressible regime is underway.

\section*{Acknowledgments}
Computing time has been provided by the Italian supercomputing center CINECA under the ISCRA C project DragBump. Parts of this work have been presented by M.Q. at the European Drag Reduction and Flow Control Meeting, Bad Herrenalb (D), Mar. 26–29 2019, and by J.B. at the XVII Euromech Turbulence Conference, Turin (I), Sept. 3-6 2019.

\section*{Declaration of Interests} 
The authors report no conflict of interest.

\bibliographystyle{jfm}

\end{document}